\documentclass[submission,copyright,creativecommons]{eptcs}
\providecommand{\event}{IMPEX  2017} % Name of the event you are submitting to 
\usepackage{breakurl}             % Not needed if you use pdflatex only. 
\usepackage{underscore}           % Only needed if you use pdflatex.

\usepackage[utf8]{inputenc}
\usepackage[T1]{fontenc}

\usepackage[noline]{algorithm2e}

\usepackage{amsfonts}
\usepackage{amsmath}
\usepackage{amssymb}
\usepackage{stmaryrd}

\usepackage{multirow}
\usepackage{url}
\usepackage{skull}
\usepackage{color}
\usepackage{placeins}
\usepackage{alltt}

\usepackage{breakurl}
\usepackage{underscore}

\usepackage{tikz}
\newcommand{\mathsc}[1]{{\normalfont\textsc{#1}}}

\newcommand{\cubiclew}{Cubicle-$\mathcal{W}$\xspace}

\newcommand{\LD}{\ensuremath{\mathcal{L_D}}\xspace}
\newcommand{\LB}{\ensuremath{\mathcal{L_B}}\xspace}
\newcommand{\LE}{\ensuremath{\mathcal{L_E}}\xspace}

\newcommand{\rrbracketu}{\ensuremath{\rrbracket_u}\xspace}
\newcommand{\rrbracketue}{\ensuremath{\rrbracket_{u_E}}\xspace}
\newcommand{\rrbracketuv}{\ensuremath{\rrbracket_{u_V}}\xspace}
\newcommand{\rrbracketg}{\ensuremath{\rrbracket_\gamma}\xspace}
\newcommand{\rrbracketge}{\ensuremath{\rrbracket_{\gamma_E}}\xspace}
\newcommand{\rrbracketgv}{\ensuremath{\rrbracket_{\gamma_V}}\xspace}

\newcommand{\D}{\ensuremath{\mathcal{D}}\xspace}
\newcommand{\M}{\ensuremath{\mathcal{M}}\xspace}
\newcommand{\Q}{\ensuremath{\mathcal{Q}}\xspace}
\newcommand{\V}{\ensuremath{\mathcal{V}}\xspace}
\renewcommand{\S}{\ensuremath{\mathcal{S}}\xspace}

\newcommand{\is}{::=}
\newcommand{\nt}[1]{\textit{#1}}
\newcommand{\kw}[1]{\textsf{#1}}
\newcommand{\vb}{\textbar\xspace}

\title{Parameterized Model Checking\\Modulo Explicit Weak Memory Models\footnote{Work supported by French ANR project PARDI (ANR-16-CE25-0006)}}

\def\titlerunning{A Backward Reachability Algorithm for
              Parameterized Systems on Weak Memory}
\def\authorrunning{S. Conchon, D. Declerck \& F. Za\"idi}

\author{Sylvain Conchon
\institute{LRI (CNRS \& Univ. Paris-Sud),\\
           Universit\'e Paris-Saclay, F-91405 Orsay}
\institute{Inria, Universit\'e Paris-Saclay, F-91120 Palaiseau}
\email{sylvain.conchon@lri.fr}
\and
David Declerck
\institute{LRI (CNRS \& Univ. Paris-Sud),\\
           Universit\'e Paris-Saclay, F-91405 Orsay}
\institute{Inria, Universit\'e Paris-Saclay, F-91120 Palaiseau}
\email{david.declerck@u-psud.fr}
\and
Fatiha Za\"idi
\institute{LRI (CNRS \& Univ. Paris-Sud),\\
           Universit\'e Paris-Saclay, F-91405 Orsay}
\email{fatiha.zaidi@lri.fr}
}

\begin{document}

\maketitle

\begin{abstract}
We present a modular framework for model checking parameterized
array-based transition systems with explicit access operations on weak
memory. Our approach extends the MCMT (Model Checking Modulo Theories)
framework of Ghilardi and Ranise~\cite{arraybased} with explicit weak
memory models. We have implemented this new framework in \cubiclew, an
extension of the Cubicle model checker. The modular architecture of
our tool allows us to change the underlying memory model seamlessly
(TSO, PSO...).  Our first experiments with a TSO-like memory model
look promising.
\end{abstract}

\section{Introduction}

% Making explicit in the Cubicle input language the weak memory model
% Param mc -> cubicle
% read X, write X ???
% -> SC (implicitly)
% modern architecture != SC

With the emerging of multi-core architectures, concurrent (or
multi-threading) programming is becoming a standard practice for
boosting the efficiency of an application. To be as efficient as
possible, concurrent programs are designed to be run for an arbitrary
number of cores. Unfortunately, in practice, the conception and
programming of such parameterized programs is error-prone and hard to
debug.

The situation is even worse if we consider that modern computer
architectures feature weak memory models in which the different
processes of a program may not perceive memory operations to happen in
the same order. For instance, under the TSO memory model, write
operations made by a process might not be immediately visible to all
other processes, while they are instantly visible to the process that
performs them. The new behaviors induced by these memory models make
it hard to design concurrent programs as one has now to take into
account both interleavings and reordering of events.

To help debugging such applications, one can use model
checking~\cite{parammc,undipcav2007,abdullatacas2007,German92,AK86,dualtso},
an efficient formal technique used for verifying safety of
parameterized concurrent programs~\cite{parammc}. For instance, one
can use Cubicle~\cite{cubicletool}, a model checker for array-based
transition systems~\cite{arraybased}, a restricted class of
parameterized systems where states are represented as logical formulas
manipulating (unbounded) arrays indexed by process identifiers.

However, the MCMT~\cite{mcmt} (Model Checking Modulo Theories)
framework underlying Cubicle \emph{implicitly} assumes a sequentially
consistent (SC) memory model: the semantics of read and write
operations is simply given by the order in which the operations are
executed, and the process actually performing the operation is
irrelevant (all processes share the same view of the memory).

In this paper, we propose an extension of MCMT~\cite{mcmt} with
\emph{explicit} read and write operations for weak memories. Our weak
memory reasoning is based on the axiomatic framework of Alglave
\textit{et al.}~\cite{herdingcats}, which describes the semantics of
different weak memory models using \emph{events} and \emph{relations}
over these events. More specifically, read and write operations on
weak variables give rise to \emph{events}, and according to the
dependencies between these events, we build a
\emph{global-happens-before} relation (\emph{ghb}) over these events,
which orders these events in a global timeline, as depicted in the
schema below.

\vspace{1em}

\begin{center}
\begin{tikzpicture}[yscale=-1, scale=1, every node/.style={scale=1}]

  \draw (0,0) node [left] {$P_1$} ;
  \draw (1,0) node [above] {$e_1$} ;
  \draw [|-] (0,0) -- ++(1,0) ;
  \draw [|-] (1,0) -- ++(4,0) ;
  \draw [dashed,->] (5,0) -- ++(1,0) ;

  \draw (0,1) node [left] {$P_2$} ;
  \draw (2,1) node [above, xshift=5] {$e_2$} ;
  \draw (4,1) node [above] {$e_6$} ;
  \draw [|-] (0,1) -- ++(2,0) ;
  \draw [|-] (2,1) -- ++(2,0) ;
  \draw [|-] (4,1) -- ++(1,0) ;
  \draw [dashed,->] (5,1) -- ++(1,0) ;

  \draw (0,1.5) node [xshift=-10,rotate=90] {...} ;

  \draw (0,2) node [left] {$P_{n-1}$} ;
  \draw (2,2) node [above,xshift=5] {$e_3$} ;
  \draw (3,2) node [above] {$e_5$} ;
  \draw (5,2) node [above] {$e_7$} ;
  \draw [|-] (0,2) -- ++(2,0) ;
  \draw [|-] (2,2) -- ++(1,0) ;
  \draw [|-] (3,2) -- ++(2,0) ;
  \draw [dashed,|->] (5,2) -- ++(1,0) ;

  \draw (0,3) node [left] {$P_n$} ;
  \draw (1,3) node [above] {$e_4$} ;
  \draw [|-] (0,3) -- ++(1,0) ;
  \draw [|-] (1,3) -- ++(4,0) ;
  \draw [dashed,->] (5,3) -- ++(1,0) ;

  \draw [->] (1,0) -- ++(1,1) node [midway, above, sloped, rotate=-90] {ghb} ;
  \draw [->] (1,0) -- ++(1,2) node [midway, below, sloped, rotate=-130] {ghb} ;
  \draw [->] (1,3) -- ++(2,-1) node [midway, above, sloped, rotate=50,yshift=-2] {ghb} ;
  \draw [->] (4,1) -- ++(1,1) node [midway, above, sloped, rotate=-90] {ghb} ;

\end{tikzpicture}
\end{center}

\vspace{1em}

In order to build this happens-before relation (\emph{ghb}), we
instrument the backward reachability analysis by (1) generating events
for read and write operations on weak memory and (2) building on the fly
a \emph{ghb} relation on this events. The coherence of this relation is
checked by an SMT solver, modulo a weak memory theory.

In the rest of the paper, we assume a TSO-like memory
model\cite{x86tsoshort}, though this framework is modular:
by changing how the \emph{ghb} relation is built, other
weak memory models can be expressed.

\FloatBarrier

\section{Preliminaries : axiomatic memory models}

Our approach relies on an axiomatic model of weak memory.
In this section, we give a brief presentation of this
kind of models, based on the formalism of
Alglave \textit{et al.}~\cite{herdingcats}.
Our presentation will be oriented towards a TSO-like model,
but other models can be expressed using the same formalism.

We consider a concurrent program $P$ composed of $n$ processes
$i_1$...$i_n$, each executing a sequence of instructions
$s_1$...$s_n$. For simplicity, we consider instructions to
be either read or write operations on weak variables.

These instructions are mapped to events, which are given a
unique event identifier. Note that when an instruction
is executed several times (for instance in loops), it
will be given a new event identifier each time it is
considered.
A read event is described by a literal of the form
\kw{Rd$_\alpha$}$(e,i)$, where $\alpha$ is the weak variable
being read, $e$ is the event identifier, and $i$ is the
identifier of the process performing the operation.
Similarly, a write event is described by a literal of the form
\kw{Wr$_\alpha$}$(e,i)$, where $\alpha$, $e$ and $i$ have
the same meaning as before. The value associated to an event $e$ on a
variable $\alpha$ is given by a function \kw{Val$_\alpha$}$(e)$.

Then, depending on the properties of these events, different
relations are built, and a \emph{global-happens-before} relation
(\emph{ghb}) is derived from them. All these relations order
the events using their unique identifiers. The set of events
together with the different relations constitutes a candidate
execution.\ If the \emph{ghb} relation is a valid partial order
(\emph{i.e.}\ is acyclic), then the execution is considered
valid.

The first relation, \emph{program order (po)}, is implied by the
program's source code. It is a total order on all events of a
process and it orders the events in the same order as the source
code. Under our TSO-like memory model, this relation allows to
derive two new relations:
\begin{itemize}
\item \emph{preserved program order (ppo)} is a partial order on the
      events of a process which represents the events that remain ordered
      under the weak memory semantics ; it is defined as the subset of
      event pairs in \emph{po} minus the write-read pairs
\item \emph{fence} indicates which write-read pairs in \emph{po}
      are separated by a fence instruction ; it allows to maintain
      the order between events that would otherwise not be ordered
      in \emph{ppo}
\end{itemize}

The next two relations depend on the actual execution of the program:
\begin{itemize}
\item \emph{coherence (co)} is a total order on all writes to the
      same variable ; it represents the order in which the writes
      are made globally visible
\item \emph{read-from (rf)} orders each write with the reads
      it provides its value to
\end{itemize}

These two relations allow to derive two new relations:
\begin{itemize}
\item \emph{from-read (fr)} indicates reads that occur before some
      write becomes globally visible ; it is defined as follows:
      $\forall e_1,e_2,e_2 \cdotp rf(e_1, e_2) \wedge
          co(e_1, e_3) \rightarrow fr(e_2, e_3)$
\item \emph{external read-from (rfe)} is defined as the subset of
      event pairs in \emph{rf} that belong to different processes
\end{itemize}

Finally, the \emph{ghb} relation is defined as the transitive
closure of the union of some of these relations:
\[ghb = (ppo\ \cup\ fence\ \cup\ co\ \cup\ rfe\ \cup\ fr)^+ \]

This relation represents the order in which events appear to
be ordered, from a global viewpoint. The key to the process of
finding a feasible execution is thus to determine a \emph{co}
and \emph{rf} relation that make the derived \emph{ghb}
relation acyclic.

Note that the axiomatic model of Alglave \textit{et al.} specifies
an auxiliary check (sequential consistency per variable), that we
do not mention here. Indeed, for TSO, the exploration strategy we
use (described in the next section) makes this check unnecessary.

\subsubsection*{Extensions for atomicity}

In order to use this framework with array-based transition
systems, that may manipulate several different variables
in a single transition, we must make some adjustments.

First, we allow several events of the same kind and by the
same process to share the same event identifier. This is
useful for instance, if we want to have two reads by a
process to occur simultaneously, without any other event
from another process interfering. Similarly, a process may
write to two variables at the same time. This means that
the writes will be made globally visible to the other
processes simultaneously. This does not require any
particular change to the model: the form of literal we use
already allows this, and building the relations considers
the events independently. For instance, let's consider three
processes $i$, $j$ and $k$, two event identifiers $e_1$ and
$e_2$, and the four events
\kw{Wr$_\alpha$}$(e_1,i)$, \kw{Wr$_\beta$}$(e_1,i)$,
\kw{Rd$_\alpha$}$(e_2,j)$ and \kw{Rd$_\beta$}$(e_3,k)$.
The two write events use the same identifier $e_1$ and belong
to the same process $i$, but write to different variables.
Then, we may have both $rf(e_1,e_2)$ and $rf(e_1,e_3)$,
even if $e_2$ and $e_3$ do not read from the same variable.

Another extension we need is that we must be able to (optionally)
specify that a read followed by a write from the same process
occur atomically, without any other event from another process
interfering. This means that, from a global point of view,
the events \emph{appear} to happen simultaneously.
For this purpose, we add a symmetric, reflexive and
transitive \textit{ghb-equal} relation, and redefine
the $ghb$ relation as follows:
\[ghb = (ppo\ \cup\ fence\ \cup\ co\ \cup\ rfe\ \cup\ fr\ \cup\ \textit{ghb-equal})^+ \setminus \textit{ghb-equal} \]

This means that \textit{ghb-equal} is only used to expand
the transitive closure of $ghb$, however events that are
\textit{ghb-equal} are removed from the actual $ghb$
relation (otherwise we wouldn't be able to tell whether
$ghb$ is acyclic).

\section{Weak Memory Array-Based Transition Systems}

In our approach, programs are described by parameterized
transition systems, \emph{i.e.} systems manipulating variables
and process-indexed arrays using guard-action transitions.
%All transitions have at least one parameter, representing
%which process performs the actions.
%Two kind of variables may be used: non-weak (or regular)
%arrays, and weak variable or arrays.
%
From the programmer's point of view, the notion of event
is irrelevant: accesses to variables and arrays are literally
understood as direct accesses. However, during our analysis,
we try to build a \emph{ghb} relation on-the-fly, hence, we
also need to be able to represent events and the relations
over these events. To comply with these different points
of view, we define two different logic languages: a description
language \LD, in which the events are implicit, and a language
\LE that makes these events explicit. We use translation functions
to translate a system from the description language \LD to the
explicit language \LE. To factor out the common parts between
these two languages, we define a base language \LB, and we have
$\LB \subset \LD$ and $\LB \subset \LE$.

\subsubsection*{Base language}

We define the base language \LB as follows:
\begin{center}
\begin{tabular}{@{}r@{\ }c@{\ }l@{}}
\nt{const, c}      & \is & constants \\
\nt{proc, i, j, k} & \is & process variables \\
\nt{x, y, z}       & \is & regular (= non-weak) arrays \\
$\alpha$, $\beta$, $\gamma$ 
                   & \is & weak variables and arrays \\
\nt{op}			   & \is & $=$ \vb $\neq$ \vb $<$ \vb $\leq$ \\
\nt{term, t}       & \is & \nt{c} \vb \nt{i} \vb x[\nt{j}] \\
\nt{atom, a}       & \is & \nt{t} \nt{op} \nt{t} \vb
                           \kw{true} \vb \kw{false} \\
\nt{literal, l}    & \is & \nt{a} \vb $\neg$\nt{a} \\
\nt{qf\_form, qff} & \is & \nt{l} \vb \nt{qff} $\wedge$ \nt{qff}
\end{tabular}
\end{center}

This language defines quantifier free formulas, which are conjunctions
of literals (or their negation). A literal is either \kw{true}, \kw{false},
or a comparison between two terms. A term is either a constant, a process
variable, or the access to a regular array cell.

\subsubsection*{Description language}

We define the description language \LD as a superset of \LB:
\begin{center}
\begin{tabular}{@{}r@{\ }c@{\ }l@{}}
\nt{term, t}       & \is & ... \vb $\alpha$
                               \vb $\alpha$[$\nt{j}$]
                               \vb $i\ @\ \alpha$
                               \vb $i\ @\ \alpha$[$\nt{j}$] \\
\nt{atom, a}       & \is & ... \vb \kw{fence}() \\
\nt{uformula, uf}  & \is & $\forall \vec{j} : proc \ldotp$ \nt{qff} \\
\nt{eformula, ef}  & \is & $\exists \vec{j} : proc \ldotp$ \nt{qff} \\
\end{tabular}
\end{center}

The description language includes the base language, and defines
new terms for accessing the weak variables.
$i\ @\ \alpha$ and $i\ @\ \alpha$[$\nt{j}$] represent
accesses to weak variables by a specific process $i$, while
$\alpha$ and $\alpha$[$\nt{j}$] do not explicitly specify
the accessing process. The context imposes which of these
two forms of access must be used. The language also defines
a \kw{fence}() predicate, which is true for some process when
its writes become globally visible to all other processes.
Finally, we define formulas prefixed by an existentially
or universally quantified process variable.

\subsubsection*{Explicit language}

We define the explicit language \LE as a superset of \LB:
\begin{center}
\begin{tabular}{@{}r@{\ }c@{\ }l@{}}
\nt{eid, e}        & \is & event identifiers \\
\nt{term, t}       & \is & ... \vb \kw{Val$_\alpha$}(\nt{e})
                               \vb \kw{Val$_\alpha$}(\nt{e}, \nt{j})\\
\nt{atom, a}       & \is & ... \vb \kw{Rd$_\alpha$}(\nt{e}, \nt{i})
                               \vb \kw{Rd$_\alpha$}(\nt{e}, \nt{i}, \nt{j}) \\
             & & \phantom{...} \vb \kw{Wr$_\alpha$}(\nt{e}, \nt{i})
                               \vb \kw{Wr$_\alpha$}(\nt{e}, \nt{i}, \nt{j}) \\
             & & \phantom{...} \vb \kw{fence}(\nt{e}, \nt{i})
%                               \vb \nt{e} $\lessdot$ \nt{e}
%                               \vb \nt{e} $\doteq$ \nt{e} \\
                               \vb \kw{ghb}(\nt{e}, \nt{e})
                               \vb \kw{ghb-equal}(\nt{e}, \nt{e}) \\
\nt{qf\_form, qff} & \is & ... \vb \nt{qff} $\vee$ \nt{qff} \\
\nt{formula, f}    & \is & \nt{qff}
                       \vb $\forall \vec{x} : type \ldotp$ \nt{f}
                       \vb $\exists \vec{x} : type \ldotp$ \nt{f} 
\end{tabular}
\end{center}

The new \kw{Rd$_\alpha$} terms allow to represent the read events,
while the \kw{Wr$_\alpha$} terms represent the write events.
The \kw{Val$_\alpha$} terms specify the value associated to events.
%\nt{e} $\lessdot$ \nt{e} and \nt{e} $\doteq$ \nt{e} allow to
\kw{ghb}(\nt{e}, \nt{e}) and \kw{ghb-equal}(\nt{e}, \nt{e}) allow to
encode the $ghb$ and \textit{ghb-equal} relations.
The \kw{fence} predicate indicates that there is a fence
before some event $e$ (it is not to be confused with the
\kw{fence} literal in \LD, although the two are related).
We also allow for more general formulas by adding disjunctions
and quantification over types other than $proc.$

Note that \kw{Val$_\alpha$}, \kw{Rd$_\alpha$} and \kw{Wr$_\alpha$}
are defined for every weak variable or array $\alpha$.
Also, \kw{Val$_\alpha$} may be considered as regular Cubicle arrays.

\subsubsection*{Convenience notations}

For simplicity, we use the following notations to represent
the different sets of variables that we often use:
\begin{itemize}
\item $X$: the set of all regular (\emph{i.e.} non-weak) arrays ($x, y...$)
\item $\hat{X}$: the set of all regular arrays ($x, y...$) and
                 event values (\kw{Val$_\alpha$}, \kw{Val$_\beta$}...)
\item $A^0$: the set of all weak variables
\item $A^1$: the set of all weak arrays
\item $A$: the set of all weak variables and arrays ($A^0\ \cup\ A^1$)
\item $A^0_t$, $A^1_t$ and $A_t$: similar to the $A^0$, $A^1$ and $A$
      sets, but restricted to the variables and arrays manipulated
      by a transition $t$
\end{itemize}
From the programmer's point of view, only $X$ and the
different $A$'s are relevant. However, the translated
formulas in \LE mainly use $\hat{X}$, since they
explicitly manipulate events and their values.

We also often use the two following notations as
shortcuts to represent common expressions:
\[\begin{array}{rcl}
\Delta(\vec{e}) & = &
  \underset{(e_1,e_2) \in \vec{e}}{\bigwedge}e_1 \neq e_2
  \hspace{5.3em} $\emph{i.e.} all elements of $\vec{e}$ are different$
\end{array}\]
\[\begin{array}{rcl}
\Diamond(\vec{e}) & = &
%  \underset{(e_1,e_2) \in \vec{e}}{\bigwedge}e_1 \doteq e_2
  \underset{(e_1,e_2) \in \vec{e}}{\bigwedge}\kw{ghb-equal}(e_1, e_2)
  \hspace{2em} $\emph{i.e.} all elements of $\vec{e}$ are \emph{ghb-equal}$
\end{array}\]

\subsubsection*{Interpretation}

The explicit language \LE is to be interpreted as follows:
\[\begin{array}{lcll}
\M[c]            & = &   \M(c) \\
\M[i]            & = &   \M(i) \\
\M[e]            & = &   \M(e) \\
\M[x[j]]   & = &   x^\M(\M[j]]) \\
\M[\kw{Val$_\alpha$}(e)]
  & = &  \kw{Val$_\alpha$}^\M(\M[e]) \\
\M[\kw{Val$_\alpha$}(e, j)]
  & = &  \kw{Val$_\alpha$}^\M(\M[e], \M[j]) \\
\M \models t_1 \ op\ t_2
  & = &  \M[t_1] \ op\ \M[t_2] \\
%\M \models e_1 \lessdot e_2
%  & = &  (\M[e_1], \M[e_2]) \in \kw{ghb}^\M \\
%\M \models e_1 \doteq e_2
%  & = &  (\M[e_1], \M[e_2]) \in \kw{ghb-equal}^\M \\
\M \models \kw{ghb}(e_1, e_2)
  & = &  (\M[e_1], \M[e_2]) \in \kw{ghb}^\M \\
\M \models \kw{ghb-equal}(e_1, e_2)
  & = &  (\M[e_1], \M[e_2]) \in \kw{ghb-equal}^\M \\
\M \models \kw{Rd$_\alpha$}(e, i)
  & = &  (\M[e], \M[i]) \in \kw{Rd$_\alpha$}^\M \\
\M \models \kw{Rd$_\alpha$}(e, i, j)
  & = &  (\M[e], \M[i], \M[j]) \in \kw{Rd$_\alpha$}^\M \\
\M \models \kw{Wr$_\alpha$}(e, i)
  & = &  (\M[e], \M[i]) \in \kw{Wr$_\alpha$}^\M \\
\M \models \kw{Wr$_\alpha$}(e, i, j)
  & = &  (\M[e], \M[i], \M[j]) \in \kw{Wr$_\alpha$}^\M \\
\M \models \kw{fence}(e, i)
  & = &  (\M[e], \M[i]) \in \kw{fence}^\M \\
\M \models \neg a
  & = &  \M \not \models a \\
\M \models \nt{qff}_1 \wedge \nt{qff}_2 
  & = &  \M \models \nt{qff}_1 $ and $ \M \models \nt{qff}_2 \\
\M \models \nt{qff}_1 \vee \nt{qff}_2 
  & = &  \M \models \nt{qff}_1 $ or $ \M \models \nt{qff}_2 \\
\M \models \forall x:type \ldotp f
  & = &  \M\{x \mapsto v\} \models f
    &  $for all $ v \in \D^{type} \\
\M \models \exists x:type \ldotp f
  & = &  \M\{x \mapsto v\} \models f
    &  $for some $ v \in \D^{type} \\
\end{array}\]

The domain of the model is partitioned according to the types $proc$,
representing the process identifiers, $eid$, representing the event
identifiers, and $val$, for the different values of variables and arrays.
We have $\D_{\M} = \D^{proc} \uplus \D^{eid} \uplus \D^{val}$.

We also define $\D^{ur} \subseteq \D^{eid}$ the subset of event
identifiers that contains only unsatisfied read events, \emph{i.e}
read events that are not connected to any write, and we have:
%
%\[\begin{array}{l}
%  \forall e_r:nc \ldotp check\_ur(e_r)
%\end{array}\]
%\[\begin{array}{rcl}
%  check\_ur(e_r) & = &
%    \forall i,j,k:proc,
%    \forall e_w:eid \ldotp e_r = e_w \vee \\
% & & \bigg( \Big(\underset{\alpha}{\bigvee}\ 
%       \kw{Rd$_{\alpha}$}(e_r, i, k) \wedge
%       \kw{Wr$_{\alpha}$}(e_w, j, k) \Big)
%         \rightarrow e_r \lessdot e_w \bigg)
%\end{array}\]
%
%Written differently:
\[\begin{array}{l}
  \forall e_r \in \D^{ur} \ldotp unsat\_read(e_r)
\end{array}\]
\[\begin{array}{rcl}
  unsat\_read(e_r) & = &
    \forall i,j,k \in \D^{proc},
    \forall e_w \in \D^{eid} \ldotp e_r = e_w \vee \\
 & &  \bigg(
        \Big(\underset{\alpha \in A^0}{\bigvee}\ 
          (e_r, i) \in \kw{Rd$_{\alpha}$} \wedge
          (e_w, j) \in \kw{Wr$_{\alpha}$} \Big)\ \vee \\
 & & \ \ \ \Big(\underset{\alpha \in A^1}{\bigvee}\ 
          (e_r, i, k) \in \kw{Rd$_{\alpha}$} \wedge
          (e_w, j, k) \in \kw{Wr$_{\alpha}$} \Big)
        \rightarrow (e_r, e_w) \in \kw{ghb} \bigg)
\end{array}\]

\subsubsection*{Initial state}

The initial state describes the constraints on regular arrays
and weak variables and arrays. It is described by a formula in
\LD, parameterized by a universally quantified process variable.
Accesses to weak variables must use the $\alpha$ or $\alpha[j]$
form ($i\ @\ \alpha$ and $i\ @\ \alpha[j]$ are not allowed).
Also, the \kw{fence()} literal may not be used.

We consider an initial state formula $I$, parameterized by
a set of regular variables X and written as follows in the
description language \LD:
\[I(X) = \forall j:proc \ldotp \mathcal{I}(j, X) \]

To obtain the equivalent formula $\tilde{I}$ in \LE, we
apply the transformation function $\llbracket \rrbracket_I$:
\[\tilde{I}(\hat{X}) = \llbracket I(X) \rrbracket_I =
  \forall i,j:proc, \forall \vec{e_\alpha}:ur \ldotp
    \underset{\alpha \in A^0}{\bigwedge}\ 
      \kw{Rd$_{\alpha}$}(e_{\alpha}, i)\ \wedge
    \underset{\alpha \in A^1}{\bigwedge}\ 
      \kw{Rd$_{\alpha}$}(e_{\alpha}, i, j)\ \wedge\ 
        \llbracket \mathcal{I}(j, X) \rrbracket_I \]

This function generates a read event for every weak variable
or array of the system, not only those actually used in $I$
(however, only those will also have a value associated to).
We have one event identifier $e_\alpha$ per weak variable or
array $\alpha$. The event identifiers are chosen in the domain
$\D^{ur}$, which restricts the events to the reads that must
take their value from the initial state. The process performing
the operation is represented by the universally quantified
variable $i$. This means that, in the initial state, any process
reading a weak variable or array will obtain the same value.
Note that since the resulting formula makes use of event value
terms (\kw{Val$_\alpha$}), the formula $\tilde{I}$ is parameterized
by the set $\hat{X}$ (while I was parameterized by $X$).

The transformation function
$\llbracket \rrbracket_I : \LD \rightarrow \LE$
is defined as follows:
\[\begin{array}{rcl@{\hspace{2em}}l}
\llbracket \nt{qff}_1 \wedge \nt{qff}_2 \rrbracket_I & = &
  \llbracket \nt{qff}_1 \rrbracket_I \wedge
    \llbracket \nt{qff}_2 \rrbracket_I \\
\llbracket \neg a \rrbracket_I & = &
  \neg \llbracket a \rrbracket_I \\
\llbracket \kw{true} \rrbracket_I & = &
  \kw{true} \\
\llbracket \kw{false} \rrbracket_I & = &
  \kw{false} \\
\llbracket t_1\ op\ t_2 \rrbracket_I & = &
  \llbracket t_1 \rrbracket_I \ op\ \llbracket t_2 \rrbracket_I \\
\llbracket \alpha \rrbracket_I & = &
  \kw{Val$_\alpha$}(e_\alpha) \\
\llbracket \alpha[j] \rrbracket_I & = &
  \kw{Val$_\alpha$}(e_\alpha, j) \\
\llbracket t \rrbracket_I & = & t &
  $when $ t \neq \alpha $ and $ t \neq \alpha[j]
\end{array}\]

\subsubsection*{States}

States represent the contraints on regular arrays, events,
events values and relations. They are not meant to be
manipulated directly by the programmer, so they are
expressed in \LE, and are parameterized by a set $\hat{X}$.

A state is represented by a formula of the following form:
\[\varphi(\hat{X}) = \exists \vec{e}:eid \ldotp
    \Delta(\vec{e}) \wedge \Phi(\vec{e}, \hat{X})\]

This means that a state uses a set of event identifiers
that are all different. The second part of the formula,
$\Phi$, is of the following form:
\[\Phi(\vec{e}, \hat{X}) = \exists \vec{j}:proc \ldotp
    \Delta(\vec{j}) \wedge \phi(\vec{e}, \vec{j}, \hat{X})\]

Similarly to events, the state uses a set of process
identifiers that are all different.
$\phi(\vec{e}, \vec{j}, \hat{X})$ is a conjunction
of literals that actually describes the constraints.

\subsection*{Bad states}

Bad states allow to describe the dangerous states of the system
in terms of constraints on regular arrays and weak variables and
arrays. They are described by a formula in \LD, parameterized
by a set of regular variables $X$. They make use of a set of
existentially quantified process variables.
Contrary to the initial state, different processes may have
different views of some weak variable, so accesses to weak
variables must use the $i\ @\ \alpha$ and $i\ @\ \alpha[j]$
form ($\alpha$ and $\alpha[j]$ are not allowed).
Also, \kw{fence()} may not be used.

We consider a bad state formula $\Theta$, written
as follows in the description language \LD:
\[\Theta(X) = \exists \vec{j}:proc \ldotp
  \Delta(\vec{j}) \wedge \vartheta(\vec{j}, X)\]

To obtain the equivalent formula $\tilde{\Theta}$ in \LE,
we apply the transformation function $\llbracket \rrbracket_u$:
\[\tilde{\Theta}(\hat{X}) =
\llbracket \Theta(X) \rrbracketu =
  \exists \vec{e}:eid \ldotp
    \Delta(\vec{e}) \wedge \Diamond(\vec{e}) \wedge
    \theta(\vec{e}, X)\]
Where:
\[\theta(\vec{e}, X) =
  \exists \vec{j}:proc \ldotp
    \Delta(\vec{j}) \wedge
    \llbracket \vartheta(\vec{j}, X)\rrbracketu^{\vec{e}}\]

The translation function transforms every weak variable or array
access to a read event, and generates one event identifier $e_i$
per process.

The translation function
$\llbracket \rrbracketu : \LD \times \vec{eid} \rightarrow \LE$
is defined as follows:

\[\begin{array}{rcl@{\hspace{2em}}l}
\llbracket \nt{qff}_1 \wedge \nt{qff}_2 \rrbracketu^{\vec{e}} & = &
  \llbracket \nt{qff}_1 \rrbracketu^{\vec{e}} \wedge
    \llbracket \nt{qff}_2 \rrbracketu^{\vec{e}} \\
\llbracket a \rrbracket_u^{\vec{e}} & = &
  \llbracket a \rrbracketue^{\vec{e}} \wedge
    \llbracket a \rrbracketuv^{\vec{e}} \\
\llbracket \neg a \rrbracket_u^{\vec{e}} & = &
  \llbracket a \rrbracketue^{\vec{e}} \wedge
    \neg \llbracket a \rrbracketuv^{\vec{e}} \\
\llbracket t_1 \ \nt{op}\ t_2 \rrbracketue^{\vec{e}} & = &
  \llbracket t_1 \rrbracketue^{\vec{e}} \wedge
    \llbracket t_2 \rrbracketue^{\vec{e}} \\
\llbracket t_1 \ \nt{op}\ t_2 \rrbracketuv^{\vec{e}} & = &
  \llbracket t_1 \rrbracketuv^{\vec{e}} \nt{op}\ 
    \llbracket t_2 \rrbracketuv^{\vec{e}} \\
\llbracket i\ @\ \alpha \rrbracketue^{\vec{e}} & = &
  \kw{Rd$_{\alpha}$}(e_i, i)      & e_i \in \vec{e} \\
\llbracket i\ @\ \alpha \rrbracketuv^{\vec{e}} & = &
  \kw{Val$_{\alpha}$}(e_i)        & e_i \in \vec{e} \\
\llbracket i\ @\ \alpha[j] \rrbracketue^{\vec{e}} & = &
  \kw{Rd$_{\alpha}$}(e_i, i, j)   & e_i \in \vec{e} \\
\llbracket i\ @\ \alpha[j] \rrbracketuv^{\vec{e}} & = &
  \kw{Val$_{\alpha}$}(e_i, j)     & e_i \in \vec{e} \\
\llbracket t \rrbracketue^{\vec{e}} & = &
  \kw{true}                       &
      $when $ t \neq i\ @\ \alpha $ and $ t \neq i\ @\ \alpha[j] \\
\llbracket t \rrbracketuv^{\vec{e}} & = &
  t                               &
      $when $ t \neq i\ @\ \alpha $ and $ t \neq i\ @\ \alpha[j]
\end{array}\]

Note that the function $\llbracket \rrbracketu$ makes
use of two sub-functions $\llbracket \rrbracketue$ and
$\llbracket \rrbracketuv$. The first one is used to build
the literals describing the events, while the second one
is used to build the event value terms.

\subsubsection*{Transitions}

Transitions describe the changes made to regular arrays
and weak variables and arrays. They are composed of a guard,
representing the conditions that must be satisfied for the transition
to be executed, and actions, which may be updates of regular
arrays or and updates of weak variables and arrays.
They are expressed in the description language \LD, with
the restriction that accesses two weak variables must
use the $\alpha$ or $\alpha[j]$ form ($i\ @\ \alpha$ and
$i\ @\ \alpha[j]$ are not allowed). Also, the \kw{fence()}
predicate may only be used in the guard.

We consider a transition $t$, written
as follows in the description language \LD:
\[\begin{array}{rcl}
t(X, X') & = &
  \exists i,\vec{j}:proc \ldotp \Delta(\vec{j}) \wedge
    \gamma(i, \vec{j}, X)\ \wedge \\ & &
  \ \underset{x \in X}{\bigwedge}
    \ x'[i] = \delta_x(i, \vec{j}, X)\ \wedge \\ & &
  \underset{{\alpha} \in A^0_t}{\bigwedge}
    \alpha' = \delta_\alpha(i, \vec{j}, X)\ \wedge \\ & &
  \underset{{\alpha} \in A^1_t}{\bigwedge}
  \ \underset{k \in \vec{j}}{\bigwedge}
    \ \alpha'[k] = \delta_\alpha(i, \vec{j}, X)
\end{array}\]

The existentially quantified process $i$ is the process
performing the actions. This also means non-weak array
terms are restricted to $x[i]$. Note that process $i$
may be equal to some process in $\vec{j}$.

To obtain the equivalent transition $\tilde{t}$ in \LE,
we apply the transformation function $\llbracket \rrbracket_t$,
which must be given two fresh event identifiers $e_r$ and $e_w$:
\[\begin{array}{rcl}
\tilde{t}(\hat{X}, \hat{X}', e_r, e_w) & = &
  \llbracket t(X, X') \rrbracket_t^{e_r,e_w}
\end{array}\]

Where:
\[\begin{array}{rcl}
\llbracket t(X, X') \rrbracket_t^{e_r,e_w} & = &
  \exists i,\vec{j}:proc \ldotp \Delta(\vec{j}) \wedge
%    e_r \neq e_w \wedge e_r \doteq e_w \wedge
    e_r \neq e_w \wedge \kw{ghb-equal}(e_r, e_w) \wedge
  \llbracket
    \gamma(i, \vec{j}, X)
  \rrbracketg^{i,e_r}\ \wedge \\ & &
  \underset{x \in X}{\bigwedge}
    \Big(
      \llbracket
        x'[i] = \delta_x(i, \vec{j}, X)
      \rrbracketg^{i,e_r} \wedge
      (\forall k. k = i\ \vee\ x'[k] = x[k])
    \Big)\ \wedge \\ & &
  \underset{{\alpha} \in A^0_t}{\bigwedge}
    \kw{Wr$_{\alpha}$}(e_w, i) \wedge
      \llbracket \kw{Val$_{\alpha}'$}(e_w) =
        \delta_\alpha(i, \vec{j}, X)
      \rrbracketg^{i,e_r}\ \wedge \\ & &
  \underset{{\alpha} \in A^1_t}{\bigwedge}
  \ \underset{k \in \vec{j}}{\bigwedge}
    \ \kw{Wr$_{\alpha}$}(e_w, i, k) \wedge
      \llbracket \kw{Val$_{\alpha}'$}(e_w, k) =
        \delta_\alpha(i, \vec{j}, X)
      \rrbracketg^{i,e_r}
\end{array}\]

This translation ensures the two event identifiers $e_r$
and $e_w$ are different, and links them in the \emph{ghb-equal}
relation. The regular array updates are extended so that all
array cells different from $i$ receive a value equals to the
previous one. The weak variable and array updates generate
write events. The translation of the guard and updates is
further delegated to the function
$\llbracket \rrbracketg : \LD \times proc \times eid \rightarrow \LE$.
\[\begin{array}{rcl@{\hspace{2em}}c}
\llbracket \nt{qff}_1 \wedge \nt{qff}_2 \rrbracketg^{i,e_r} & = &
  \llbracket \nt{qff}_1 \rrbracketg^{i,e_r} \wedge
    \llbracket \nt{qff}_2 \rrbracketg^{i,e_r} \\
\llbracket a \rrbracketg^{i,e_r} & = &
  \llbracket a \rrbracketge^{i,e_r} \wedge
    \llbracket a \rrbracketgv^{i,e_r} \\
\llbracket \neg a \rrbracketg^{i,e_r} & = &
  \llbracket a \rrbracketge^{i,e_r} \wedge
    \neg \llbracket a \rrbracketgv^{i,e_r} \\
\llbracket \kw{fence}() \rrbracketge^{i,e_r} & = & \kw{fence}(e_r, i) \\
\llbracket \kw{fence}() \rrbracketgv^{i,e_r} & = & \kw{true} \\
\llbracket t_1 \nt{ op } t_2 \rrbracketge^{i,e_r} & = &
  \llbracket t_1 \rrbracketge^{i,e_r} \wedge
    \llbracket t_2 \rrbracketge^{i,e_r} \\
\llbracket t_1 \nt{ op } t_2 \rrbracketgv^{i,e_r} & = &
  \llbracket t_1 \rrbracketgv^{i,e_r} \nt{ op }
    \llbracket t_2 \rrbracketgv^{i,e_r} \\
\llbracket \alpha \rrbracketge^{i,e_r} & = &
  \kw{Rd$_{\alpha}$}(e_r, i) \\
\llbracket \alpha \rrbracketgv^{i,e_r} & = &
  \kw{Val$_{\alpha}$}(e_r, i) \\
\llbracket \alpha[j] \rrbracketge^{i,e_r} & = &
  \kw{Rd$_{\alpha}$}(e_r, i, j) \\
\llbracket \alpha[j] \rrbracketgv^{i,e_r} & = &
  \kw{Val$_{\alpha}$}(e_r, i) \\
\llbracket t \rrbracketge^{i,e_r} & = & \kw{true} &
      $when $ t \neq \alpha $ and $ t \neq \alpha[j] \\
\llbracket t \rrbracketgv^{i,e_r} & = & t         &
      $when $ t \neq \alpha $ and $ t \neq \alpha[j] \\
\end{array}\]

\section{Backward reachability}
\label{sec:bwd}

Our approach relies on a rather classical backward reachability
algorithm (function \texttt{BWD} below), whose objective is to
check whether there is a possible path from the initial state to
the dangerous states. However, the pre-image computation has
been extended to produce events and relations, according to
our weak memory semantics.
The algorithm takes as input a transition system $S = (Q, I, \tau)$ and
a cube $\Theta$, where $I$ is a formula describing the initial states of
the system, $\tau$ is the set of all transitions, and $\Theta$ a cube
describing the dangerous states. It maintains a set of visited states
$\mathcal{V}$ and a working queue of cubes $\mathcal{Q}$.

\LinesNumbered
\begin{algorithm}[h]
\SetKwProg{Fn}{function}{ {\normalfont :} begin}{}
\Fn{$\mathsc{Bwd}(\S, \Theta)$}{
	$\V := \emptyset$\;
	$\mathit{push}(\Q, \llbracket \Theta \rrbracketu)$\;
	\While{$\mathit{not\_empty}(\Q)$}{
		$\varphi := \mathit{pop}(\Q)$\;
		\If{$\mathsf{\varphi \wedge
		      \mathit{\llbracket I \rrbracket_I} \ satisfiable}$}{
		    \Return{unsafe}
		}
		\ElseIf{$\varphi \nvDash \V$}{
			$\V := \V \cup \{ \varphi \}$\;
			$\mathit{push}(\Q, \mathsc{Pre}_\tau(\varphi))$;
		}
	}
	\Return{safe}
}
\end{algorithm}

\FloatBarrier

We assume the fomulas describing states to be of the form:
\[\varphi(\hat{X}) = \exists \vec{e}:eid \ldotp
    \Delta(\vec{e}) \wedge \Phi(\vec{e}, \hat{X})\]

The pre-image of a formula $\varphi$ with respect to the set of
transitions $\tau$ (line 11) is given by:
\[\mathsc{Pre}_\tau(\varphi)(\hat{X}) =
  \underset{t \in \tau}{\bigvee}\ \mathsc{Pre}_t(\varphi)(\hat{X})\]

The pre-image of a formula $\varphi$ with respect to a single
transition $t$ is given by:
\[\begin{array}{rcl}
\mathsc{Pre}_t(\varphi)(\hat{X}) & = &
  \exists \hat{X}', \exists e_r, e_w, \vec{e}:eid \ldotp
    \Delta(e_r \cdotp e_w \cdotp \vec{e})\ \wedge \\ & &
  \llbracket t(X,X') \rrbracket_t^{e_r,e_w} \wedge
    \Phi(\vec{e}, \hat{X}')\ \wedge \\ & &
  extend\_ghb(e_r, e_w, \vec{e}) \wedge rffr(X, X', e_w, \vec{e})
\end{array}\]

This pre-image generates two new event identifiers $e_r$ and $e_w$.
We ensure these identifiers to be different from those in $\varphi$
using the expression $\Delta(e_r \cdotp e_w \cdotp \vec{e})$.
These identifiers are given as parameters to the translation
function $\llbracket \rrbracket_t$, which ensures the transition
only manipulates new events.

The $rffr$ function determines whether the new writes $e_w$ from
the transition $t$ satisfy the compatible unsatisfied reads in
$\Phi(\vec{e}, \hat{X}')$, and if so orders them appropriately
in the \emph{ghb} relation. A read and a write are compatible
if they refer to the same variable or the same array with the
same parameters (the actual value associated to the events
is irrelevant to this notion of compatibility).
For each unsatisfied read in $\vec{e}$, either:
\begin{itemize}
\item there is a compatible write $e_w$ from the \emph{same} process,
      so the read MUST take its value from this write
\item there is a compatible write $e_w$ from a \emph{different} process,
      in this case the read MAY take its value from this write ;
      if it does, $e_w$ is before $e_r$ in $ghb$,
      otherwise $e_r$ is before $e_w$ in $ghb$
\item there is no compatible write ; the read remains unsatisfied
\end{itemize}
This implies the following logical definition of $rffr$:
%
%\[\begin{array}{rcl}
%rffr(X, X', e_w, \vec{e}) & = &
%  \underset{{\alpha} \in A}{\bigwedge}\ 
%   \underset{e_r \in \vec{e}}{\bigwedge}
%    \bigg( \forall i, k:proc \ldotp
%      unsat\_read_\alpha(e_r, i, k, \vec{e}) \rightarrow \\
%
%& & \ \ \Big( \kw{Wr$_{\alpha}$}(e_w, i, k)\ \wedge \\
%& & \ \ \ \ \kw{Val$_{\alpha}$'}(e_r, k) =
%              \kw{Val$_{\alpha}$'}(e_w, k)
%        \Big)\ \veebar \\
%
%& & \ \ \Big( \exists j:proc \ldotp j \neq i \wedge
%          \kw{Wr$_{\alpha}$}(e_w, j, k)\ \wedge \\
%& & \ \ \ \ (\kw{Val$_{\alpha}$'}(e_r, k) =
%               \kw{Val$_{\alpha}$'}(e_w, k) \wedge
%               e_w \lessdot e_r \ \vee \\
%& & \ \ \ \ \ \kw{Val$_{\alpha}$'}(e_r, k) =
%               \kw{Val$_{\alpha}$}(e_r, k) \wedge
%               e_r \lessdot e_w)
%        \Big)\ \veebar \\
%
%& & \ \ \Big( (\not \exists j:proc \ldotp
%          \kw{Wr$_{\alpha}$}(e_w, j, k))\ \wedge \\
%& & \ \ \ \ \kw{Val$_{\alpha}$'}(e_r, k) =
%              \kw{Val$_{\alpha}$}(e_r, k)
%        \Big)
%    \bigg)
%\end{array}\]
%
\[\begin{array}{rcl@{}l}
rffr(X, X', e_w, \vec{e}) & = &
  \underset{{\alpha} \in A}{\bigwedge}\ 
   \underset{e_r \in \vec{e}}{\bigwedge}
    \Big( & \exists i, k:proc \ldotp
      unsat\_read_\alpha(e_r, i, k, \vec{e}) \rightarrow \\
& & &  internal\_rffr_\alpha(X,X',e_r,e_w,i,k)\ \veebar \\
& & &  external\_rffr_\alpha(X,X',e_r,e_w,i,k)\ \veebar \\
& & &  no\_rffr_\alpha(X,X',e_r,e_w,k) \Big)
\end{array}\]
\[\begin{array}{rcl}
internal\_rffr_\alpha(X, X', e_r, e_w, i, k) & = &
  \kw{Wr$_{\alpha}$}(e_w, i, k) \wedge
    \kw{Val$_{\alpha}$'}(e_r, k) =
      \kw{Val$_{\alpha}$'}(e_w, k)
\end{array}\]
\[\begin{array}{rcl}
external\_rffr_\alpha(X, X', e_r, e_w, i, k) & = &
  \exists j:proc \ldotp j \neq i \wedge
    \kw{Wr$_{\alpha}$}(e_w, j, k)\ \wedge \\
& & (\kw{Val$_{\alpha}$'}(e_r, k) =
               \kw{Val$_{\alpha}$'}(e_w, k) \wedge
%               e_w \lessdot e_r \ \vee \\
                \kw{ghb}(e_w, e_r) \ \vee \\
& & \ \ \kw{Val$_{\alpha}$'}(e_r, k) =
               \kw{Val$_{\alpha}$}(e_r, k) \wedge
%               e_r \lessdot e_w)
                \kw{ghb}(e_r, e_w)
\end{array}\]
\[\begin{array}{rcl}
no\_rffr_\alpha(X, X', e_r, e_w, i, k) & = &
  (\not \exists j:proc \ldotp
    \kw{Wr$_{\alpha}$}(e_w, j, k)) \wedge
      \kw{Val$_{\alpha}$'}(e_r, k) =
        \kw{Val$_{\alpha}$}(e_r, k)
\end{array}\]

$rffr$ also relies on the $unsat\_read_\alpha$ function, that
allows to determine if some read event $e_r$ is not satisfied by
any write event in $\vec{e}$ for some weak variable $\alpha$:
\[\begin{array}{rcl}
unsat\_read_\alpha(e_r, i, k, \vec{e}) & = &
  \kw{Rd$_{\alpha}$}(e_r, i, k) \wedge
     \underset{e_w \in \vec{e}}{\bigwedge}
        \bigg( e_r = e_w \vee \Big( \forall j:proc \ldotp
          \kw{Wr$_{\alpha}$}(e_w, j, k) \rightarrow
%            e_r \lessdot e_w \Big) \bigg)
             \kw{ghb}(e_r, e_w) \Big) \bigg)
\end{array}\]

The \textit{extend\_ghb} function extends the \emph{ghb} relation by
%adding literals of the form $e_1 \lessdot e_2$ and $e_1 \doteq e_2$ in
adding literals of the form \kw{ghb}(e_1, e_2) and \kw{ghb-equal}(e_1, e_2) in
the formula, according to the dependencies between the new events
$e_r$ and $e_w$ and old events in $\vec{e}$.
\[\begin{array}{rcl}
extend\_ghb(e_r, e_w, \vec{e}) & = &
    ppo(e_r \cdotp e_w, \vec{e}) \wedge fence(e_w, \vec{e}) \wedge
    co(e_w, \vec{e}) \wedge fr(e_r, \vec{e})
\end{array}\]

$fr$ adds $ghb$ pairs between the new read(s)
$e_r$ and the old writes.
%
%\[\begin{array}{rcl}
%fr(e_r, \vec{e}) & = &
%  \underset{\alpha \in A}{\bigwedge}\ 
%    \underset{e_w \in \vec{e}}{\bigwedge}
%    \Big( \forall i, j, k:proc \ldotp
%       \kw{Rd$_{\alpha}$}(e_r, i, k) \wedge
%          \kw{Wr$_{\alpha}$}(e_w, j, k) \rightarrow
%            e_r \lessdot e_w \Big) \\
%\end{array}\]
%
\[\begin{array}{rcl}
fr(e_r, \vec{e}) & = &
  \underset{\alpha \in A}{\bigwedge}\ 
    \underset{e_w \in \vec{e}}{\bigwedge}
    \Big( \exists k:proc \ldotp
      read\_on_\alpha(e_r, k) \wedge
      write\_on_\alpha(e_w, k)
%      \rightarrow e_r \lessdot e_w \Big)
      \rightarrow \kw{ghb}(e_r, e_w) \Big)
\end{array}\]

$co$ adds $ghb$ pairs between the new write(s)
$e_w$ and the old writes.
%
%\[\begin{array}{rcl}
%co(e_{w_1}, \vec{e}) & = &
%  \underset{\alpha \in A}{\bigwedge}\ 
%    \underset{e_{w_2} \in \vec{e}}{\bigwedge}
%    \Big( \forall i, j, k:proc \ldotp
%       \kw{Wr$_{\alpha}$}(e_{w_1}, i, k) \wedge
%          \kw{Wr$_{\alpha}$}(e_{w_2}, j, k) \rightarrow
%            e_{w_1} \lessdot e_{w_2} \Big) \\
%\end{array}\]
%
\[\begin{array}{rcl}
co(e_{w_1}, \vec{e}) & = &
  \underset{\alpha \in A}{\bigwedge}\ 
    \underset{e_{w_2} \in \vec{e}}{\bigwedge}
    \Big( \exists k:proc \ldotp
      write\_on_\alpha(e_{w_1}, k) \wedge
      write\_on_\alpha(e_{w_2}, k)
%      \rightarrow e_{w_1} \lessdot e_{w_2} \Big)
      \rightarrow \kw{ghb}(e_{w_1}, e_{w_2} \Big)
\end{array}\]

$fence$ adds $ghb$ pairs between the new writes(s) $e_w$ and the
subsequent reads by the same process separated by a fence predicate.
%
%\[\begin{array}{rcl}
%fence(e_w, \vec{e}) & = &
%  \underset{e_r \in \vec{e}}{\bigwedge}
%    \Bigg( \forall i, k_1, k_2:proc \ldotp
%      \kw{fence}(e_r, i)\ \wedge \\
%& & \ \ \bigg( \underset{\alpha \in A}{\bigvee}
%          \ \kw{Wr$_{\alpha}$}(e_w, i, k_1) \bigg) \wedge
%        \bigg( \underset{\alpha \in A}{\bigvee}
%          \ \kw{Rd$_{\alpha}$}(e_r, i, k_2) \bigg)
%      \rightarrow e_w \lessdot e_r \Bigg)
%\end{array}\]
%
\[\begin{array}{rcl}
fence(e_w, \vec{e}) & = &
  \underset{e_r \in \vec{e}}{\bigwedge}
    \Big( \exists i:proc \ldotp
      \kw{fence}(e_r, i) \wedge
      write\_by(e_w, i) \wedge
      read\_by(e_r, i)
%      \rightarrow e_w \lessdot e_r \Big)
      \rightarrow \kw{ghb}(e_w, e_r) \Big)
\end{array}\]

$ppo$ adds $ghb$ pairs between the new events $e_a$ and the
subsequent events $e_b$ by the same process if they are
read-read, read-write or write-write pairs.
%
%
%\[\begin{array}{rcl}
%ppo(\vec{e_a}, \vec{e_b}) & = &
%  \underset{e_1 \in \vec{e_a}}{\bigwedge}\ 
%    \underset{e_2 \in \vec{e_b}}{\bigwedge}
%    \Bigg( \forall i, k_1, k_2:proc \ldotp \\
%& & \bigg( \Big(\underset{\alpha \in A}{\bigvee}
%             \ \kw{Rd$_{\alpha}$}(e_1, i, k_1) \Big) \wedge
%           \Big(\underset{\alpha \in A}{\bigvee}
%             \ \kw{Rd$_{\alpha}$}(e_2, i, k_2) \Big)
%           \rightarrow e_1 \lessdot e_2 \bigg)\ \wedge \\
%& & \bigg( \Big(\underset{\alpha \in A}{\bigvee}
%             \ \kw{Rd$_{\alpha}$}(e_1, i, k_1) \Big) \wedge
%           \Big(\underset{\alpha \in A}{\bigvee}
%             \ \kw{Wr$_{\alpha}$}(e_2, i, k_2) \Big)
%           \rightarrow e_1 \lessdot e_2 \bigg)\ \wedge \\
%& & \bigg( \Big(\underset{\alpha \in A}{\bigvee}
%             \ \kw{Wr$_{\alpha}$}(e_1, i, k_1) \Big) \wedge
%           \Big(\underset{\alpha \in A}{\bigvee}
%             \ \kw{Wr$_{\alpha}$}(e_2, i, k_2) \Big)
%           \rightarrow e_1 \lessdot e_2 \bigg)
%    \Bigg)
%\end{array}\]
%
%
%
%
\[\begin{array}{rcl}
ppo(\vec{e_a}, \vec{e_b}) & = &
  \underset{e_1 \in \vec{e_a}}{\bigwedge}\ 
    \underset{e_2 \in \vec{e_b}}{\bigwedge}
    \Big( ppo\_RR(e_1, e_2) \wedge
          ppo\_RW(e_1, e_2) \wedge
          ppo\_WW(e_1, e_2) \Big)
\end{array}\]
\[\begin{array}{rcl}
ppo\_RR(e_1, e_2) & = & \exists i:proc \ldotp
   read\_by(e_1, i)\ \wedge\ read\_by(e_2, i)
%      \rightarrow e_1 \lessdot e_2
      \rightarrow \kw{ghb}(e_1, e_2)
\end{array}\]
\[\begin{array}{rcl}
ppo\_RW(e_1, e_2) & = & \exists i:proc \ldotp
   read\_by(e_1, i)\ \wedge\ write\_by(e_2, i)
%      \rightarrow e_1 \lessdot e_2
      \rightarrow \kw{ghb}(e_1, e_2)
\end{array}\]
\[\begin{array}{rcl}
ppo\_WW(e_1, e_2) & = & \exists i:proc \ldotp
   write\_by(e_1, i)\ \wedge\ write\_by(e_2, i)
%      \rightarrow e_1 \lessdot e_2
      \rightarrow \kw{ghb}(e_1, e_2)
\end{array}\]

The following $read\_on_\alpha$ and $write\_on_\alpha$ predicates
allow to easily check whether some event identifier $e$ corresponds
to a read or a write event on a specific weak variable $\alpha$
or array cell $\alpha[k]$.
\[\begin{array}{rcl}
read\_on_\alpha(e, k) & = & \exists i:proc \ldotp
  \kw{Rd$_{\alpha}$}(e, i) \vee \kw{Rd$_{\alpha}$}(e, i, k)
\end{array}\]
\[\begin{array}{rcl}
write\_on_\alpha(e, k) & = & \exists i:proc \ldotp
  \kw{Wr$_{\alpha}$}(e, i) \vee \kw{Wr$_{\alpha}$}(e, i, k)
\end{array}\]

Similarly, the $read\_by$ and $write\_by$ predicates allow to
check whether some event identifier $e$ corresponds to a read
or a write event performed by a specific process $i$.
\[\begin{array}{rcl}
read\_by(e, i) & = & \exists k:proc \ldotp
  \underset{\alpha \in A^0}{\bigvee} \ \kw{Rd$_{\alpha}$}(e, i) \vee
  \underset{\alpha \in A^1}{\bigvee} \ \kw{Rd$_{\alpha}$}(e, i, j)
\end{array}\]
\[\begin{array}{rcl}
write\_by(e, i) & = & \exists k:proc \ldotp
  \underset{\alpha \in A^0}{\bigvee} \ \kw{Wr$_{\alpha}$}(e, i) \vee
  \underset{\alpha \in A^1}{\bigvee} \ \kw{Wr$_{\alpha}$}(e, i, j)
\end{array}\]

%How the new events connect to the old ones depends on several memory model
%specific rules, as explained in~\cite{herdingcats}. Under TSO, each new
%write in $\W$ is \emph{ghb}-before the globally first write to the same
%variable in $\FWv$ (there is a total order on all writes to the same
%variable) and the first write of the same process in $\FWt$. New reads
%in $\R$ are \emph{ghb}-before the first reads and writes of the same
%process in $\FRt$ and $\FWt$. Writes in a combination $c$ are (obviously)
%\emph{ghb}-before the reads they provide their value to. Finally, all reads
%and writes in $\R\ \cup\ \W$ are \emph{ghb}-equal (\emph{i.e.} they happen
%simultaneously in the global timeline). It is to be noted that new writes
%in $\W$ are \emph{not} \emph{ghb}-before the first reads of their respective
%processes: indeed, because of write buffering, a write may reach memory
%after a subsequent read, thus, this pair of events does not participate
%in the \emph{ghb} relation. Also note that, at any time, if we build a
%state containing an invalid (cyclic) \emph{ghb} relation, then this state
%is discarded.

\section{Example}

We illustrate our backward reachability algorithm through a simple
example. We consider a simple parameterized mutual exclusion
algorithm, where each process executes the automaton below.

\medskip
\begin{center}
\includegraphics[scale=0.7]{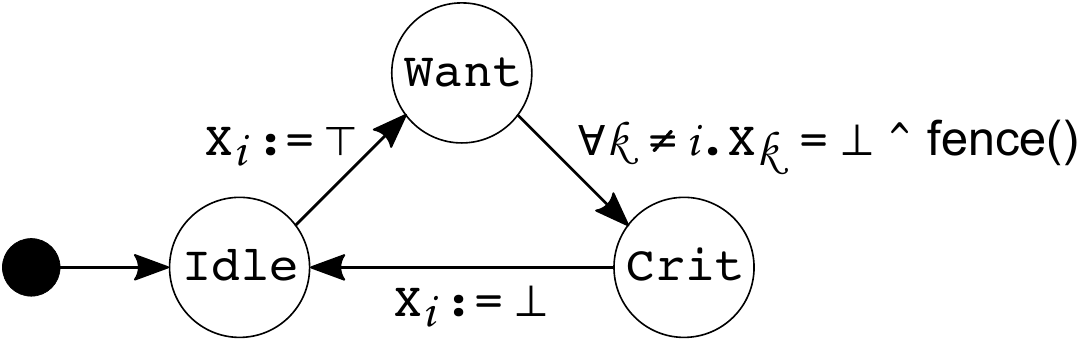}
\end{center}
\medskip

\noindent
This automaton is trivially encoded into the following transition
system, where the current state is represented by a regular array
\texttt{PC}, and the shared variables \texttt{X}$_i$ are expressed
as a weak array \texttt{X}.
\[t\_req = \exists i:proc \ldotp
  PC[i] = Idle \wedge
  PC'[i] = Want \wedge X'[i] = True\]
\[t\_enter = \exists i:proc \ldotp
  PC[i] = Want \wedge \kw{fence}() \wedge
    (\forall k. k = i \vee X[i] = False) \wedge
  PC'[i] = Crit\]
\[t\_exit = \exists i:proc \ldotp
  PC[i] = Crit \wedge
  PC'[i] = Idle \wedge X'[i] = False\]
The initial state is simply defined as:
\[I = \forall j:proc \ldotp
  PC[j] = Idle \wedge X[j] = False\]
The property we would like to check is that no pair of processes
$i$ and $j$ can be in state \texttt{Crit} at the same time,
\emph{i.e.} that the following formula never holds:
\[\Theta = \exists i,j:proc \ldotp
    i \neq j \wedge PC[i] = Crit \wedge PC[j] = Crit\]

By applying the translation functions on the system above, we
obtain the following event-explicit transition system:
\[\tilde{I} = \forall i,j:proc, \forall e_X:ur \ldotp
  PC[j] = Idle \wedge Rd_X(e_X,i,j) \wedge Val_X(e_X,j) = False\]
\[\tilde{\Theta} = \exists i,j:proc \ldotp
    i \neq j \wedge PC[i] = Crit \wedge PC[j] = Crit\]
\[\begin{array}{rcl}
  \tilde{t}\_req(e_r, e_w) & = & \exists i:proc \ldotp
%  e_r \neq e_w \wedge e_r \doteq e_w\ \wedge \\ & &
  e_r \neq e_w \wedge \kw{ghb-equal}(e_r, e_w)\ \wedge \\ & &
  PC[i] = Idle\ \wedge \\ & &
  PC'[i] = Want \wedge
  Wr_X(e_w,i,i) \wedge Val_X'(e_w,i) = True
\end{array}\]
\[\begin{array}{rcl}
  \tilde{t}\_enter(e_r, e_w) & = & \exists i:proc \ldotp
%  e_r \neq e_w \wedge e_r \doteq e_w \wedge\ \\ & &
  e_r \neq e_w \wedge \kw{ghb-equal}(e_r, e_w)\ \wedge \\ & &
  PC[i] = Want \wedge \kw{fence}(e_r,i)\ \wedge \\ & &
  (\forall k. k = i \vee Rd_X(e_r,i,k) \wedge
    Val_X(e_r,i) = False)\ \wedge \\ & &
  PC'[i] = Crit
\end{array}\]
\[\begin{array}{rcl}
  \tilde{t}\_exit(e_r, e_w) & = & \exists i:proc \ldotp
%  e_r \neq e_w \wedge e_r \doteq e_w \wedge\ \\ & &
  e_r \neq e_w \wedge \kw{ghb-equal}(e_r, e_w)\ \wedge \\ & &
  PC[i] = Crit \wedge\ \\ & &
  PC'[i] = Idle \wedge
  Wr_X(e_w,i,i) \wedge Val_X'(e_w,i) = False
\end{array}\]

The graph below gives a possible exploration of the system's state
space by our algorithm. We start by node 1, which represents the
formula describing the dangerous states $\tilde{\Theta}$.
Then, each node represents the result of a pre-image computation
by an instance of a transition. The edges are labeled with the
transition name and the process identifier we used to instantiate
the existentially quantified process variable in the transition.
Remark that formulas in the graph's nodes are implicitly existentially
quantified and that a process identifier $i$ is written $\texttt{\#}_i$.
Also, to avoid cluttering the graph, we omit event identifiers in the
edge labels, we remove the unused event identifiers from the nodes
(they do not contribute to $ghb$), and we assume all event
identifiers to be different.

\begin{center}
\includegraphics[scale=0.7]{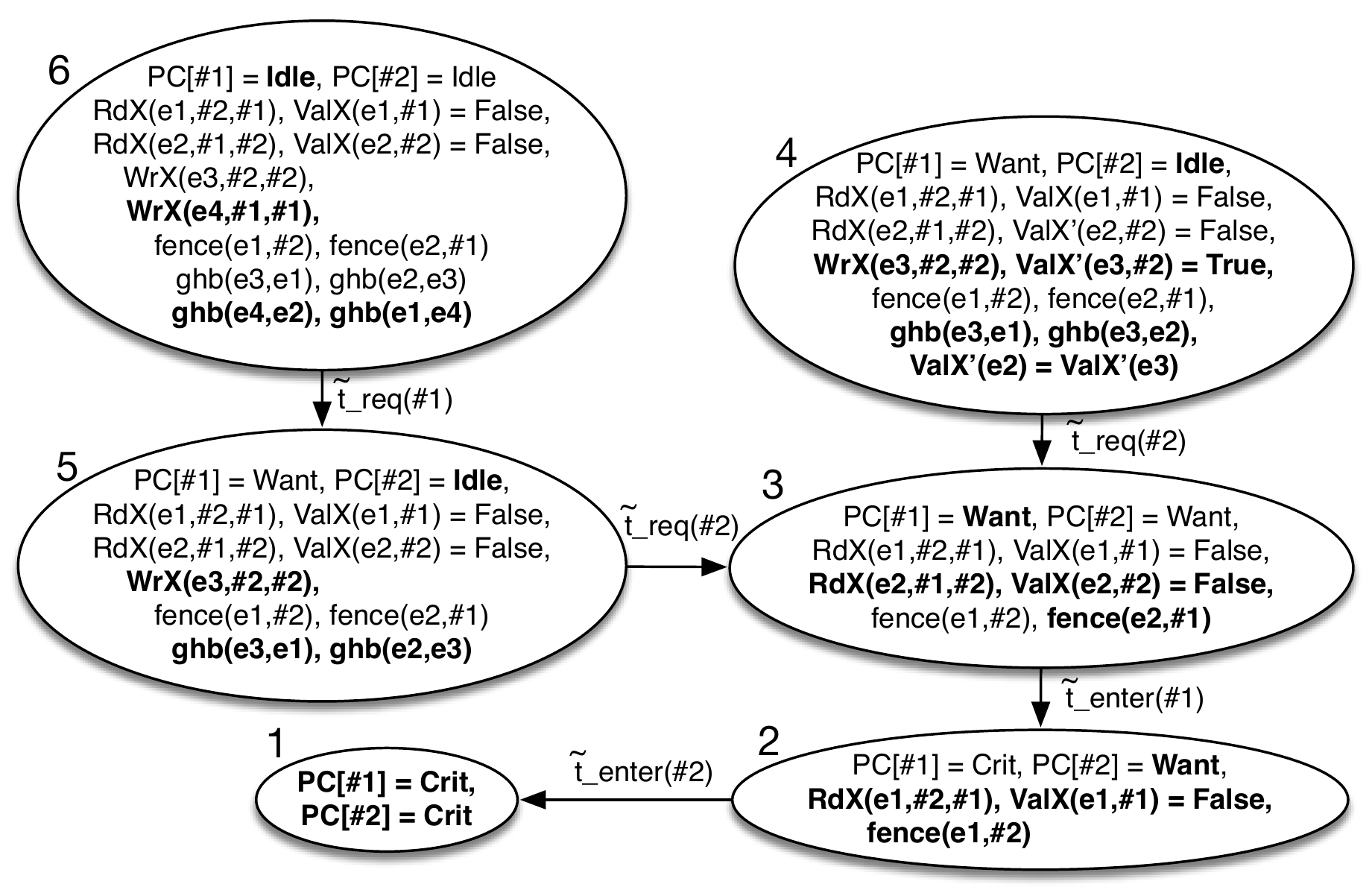}
\end{center}

We focus on node 3 which results from the pre-image of node 1 by
\texttt{t\_enter(\#2)} then \texttt{t\_enter(\#1)}. In this state,
both processes have read \texttt{False} in \texttt{X} (events
\texttt{$e_1$} and \texttt{$e_2$}). Also, since there is a memory
barrier in \texttt{t\_enter}, both reads are associated to a
\texttt{fence} literal. The pre-image of node 3 by \texttt{t\_req(\#2)}
introduces a new write event \texttt{WrX(e3,\#2,\#2)} with an
associated value \texttt{ValX'(e3,\#2) = True}.
Since there is a memory barrier \texttt{fence(e1,\#2)} on \texttt{e1}
by the same process \texttt{\#2}, the $extend\_ghb$ predicate
causes \texttt{ghb(e3,e1)} to be added to the formula.
Now, the $rffr$ predicate dictates that this new write event
may or may not satisfy the read \texttt{e2}, so we must consider
both cases (node 4 and 5).

In node 4, we consider the case where the write event
\texttt{e3} satisfies the read event \texttt{e2}.
As prescribed by the $external\_rffr$ predicate, the
equality \texttt{ValX'(e2,\#2) = ValX'(e3,\#2)} is added
to the formula, which obviously makes it inconsistent.
In node 5, the write event \texttt{e3} does not satisfy the read
event \texttt{e2}, so the value \texttt{ValX'(e3,\#2)} is discarded
and \texttt{ghb(e2,e3)} is added to the formula, as indicated
by the $no\_rffr$ predicate.
Similarly, the pre-image of node 5 by \texttt{t\_req(\#1)} yields
the formula in state 6 where the new write event \texttt{e4} does
not satisfy the read event \texttt{e1}. Now, the $ghb$ relation
is not a valid partial order as the sequence
$\texttt{ghb}(\texttt{e2},\texttt{e3}),
\texttt{ghb}(\texttt{e3},\texttt{e1}),
\texttt{ghb}(\texttt{e1},\texttt{e4}),
\texttt{ghb}(\texttt{e4},\texttt{e2})$
forms a cyclic relation. Therefore, this state is discarded
and the program is declared \emph{safe}.

Remark that if we removed the fence predicate in \texttt{t\_enter},
then we would only have $\texttt{ghb}(\texttt{e3},\texttt{e1}),$
$\texttt{ghb}(\texttt{e4},\texttt{e2})$ in state 6, which is a valid
partial order relation, so the formula would intersect with the
initial state and the program would be \emph{unsafe}.

\section{Implementation}

We have implemented this framework in \cubiclew~\cite{cubiclew} and
used it to check the correctness of several parameterized concurrent
algorithms on a TSO-like model whose source codes are available on the
tool's webpage.

Table \ref{tb:benchs} gives for each benchmark the number of non-weak
arrays, the number of weak variables, the number of weak arrays, the
number of transitions and the running time. The S/US next to the
benchmark name indicates whether the algorithm is correct or
not. Incorrect algorithm have a second version that was fixed by using
\kw{fence} predicates.

\begin{table}[h]
\centering
  \begin{tabular}{|l|c|c|c|c|c|c|}
    \hline
    \multicolumn{2}{|c|}{Case study}
      & Arrays & Weak Var. & Weak Arr. & Transitions & Time \\
    \hline
    \texttt{naive mutex} & US & 1 & 0 & 1 & 4 & 0.04s \\
    \hline
    \texttt{naive mutex} & S & 1 & 0 & 1 & 4 & 0.30s \\
    \hline
    \texttt{lamport} & US & 1 & 4 & 0 & 8 & 0.10s \\
    \hline
    \texttt{lamport} & S & 1 & 4 & 0 & 8 & 0.60s \\
    \hline
    \texttt{spinlock} & S & 1 & 1 & 0 & 6 & 0.07s \\
    \hline
    \texttt{sense reversing barrier} & S & 1 & 0 & 1 & 4 & 0.06s \\
    \hline
    \texttt{arbiter v1} & S & 2 & 1 & 1 & 7 & 0.18s \\
    \hline
    \texttt{arbiter v2} & S & 2 & 0 & 2 & 8 & 13.5s \\
    \hline
    \texttt{two phase commit} & S & 1 & 1 & 1 & 5 & 54.1s \\
    \hline
  \end{tabular}
\medskip
\caption{Performance of \cubiclew}
\label{tb:benchs}
\end{table}

%\FloatBarrier

The results shown in this table look promising.  Although these
benchmarks are of modest size, they are already consider as very
challenging for state-of-the-art model checkers for weak memories as
they combined parametricity, concurrency and non trivial used of weak
memories.

\section{Conclusion \& Future Work}

We have presented in this paper an extension of Model Checking Modulo
Theories (MCMT) for model checking parameterized transitions with
\emph{explicit} read and write operations on weak memories.

The core of our procedure is a backward reachability algorithm
combined with an SMT axiomatic model for reasoning about weak memory.
The explicit relaxed consistency model underlying our framework is
similar to x86-TSO where the effect of a store operation by a process
is delayed (due to a store buffering) to all processes.

We have implemented this framework in Cubicle-$\mathcal{W}$, a
conservative extension of the Cubicle model checker. Our first
experiments show that our implementation is expressive and efficient
enough to prove safety of concurrent algorithms, for an arbitrary
number of processes, ranging from mutual exclusion to synchronization
barriers.

Immediate future work includes the support for other models, such as
PSO. We are also working on the extension of Cubicle's invariant
generation mechanism for weak memory. With this mechanism, we should
gain in efficiency and be able to tackle even more complex programs.
Finally, we also plan to investigate how our framework for weak
memories could be extended to reason about programs communicating via
channels.

%\nocite{*}
\bibliographystyle{eptcs}

\begin{thebibliography}{10}
\providecommand{\bibitemdeclare}[2]{}
\providecommand{\surnamestart}{}
\providecommand{\surnameend}{}
\providecommand{\urlprefix}{Available at }
\providecommand{\url}[1]{\texttt{#1}}
\providecommand{\href}[2]{\texttt{#2}}
\providecommand{\urlalt}[2]{\href{#1}{#2}}
\providecommand{\doi}[1]{doi:\urlalt{http://dx.doi.org/#1}{#1}}
\providecommand{\bibinfo}[2]{#2}

\bibitemdeclare{misc}{cubiclew}
\bibitem{cubiclew}
\bibinfo{note}{\url{http://cubicle.lri.fr/cubiclew/}}.

\bibitemdeclare{inproceedings}{dualtso}
\bibitem{dualtso}
\bibinfo{author}{Parosh~Aziz \surnamestart Abdulla\surnameend},
  \bibinfo{author}{Mohamed~Faouzi \surnamestart Atig\surnameend},
  \bibinfo{author}{Ahmed \surnamestart Bouajjani\surnameend} \&
  \bibinfo{author}{Tuan~Phong \surnamestart Ngo\surnameend}
  (\bibinfo{year}{2016}): \emph{\bibinfo{title}{The Benefits of Duality in
  Verifying Concurrent Programs under {TSO}}}.
\newblock In: {\sl \bibinfo{booktitle}{27th International Conference on
  Concurrency Theory, {CONCUR} 2016, August 23-26, 2016, Qu{\'{e}}bec City,
  Canada}}, pp. \bibinfo{pages}{5:1--5:15}, \doi{10.4230/LIPIcs.CONCUR.2016.5}.

\bibitemdeclare{inproceedings}{abdullatacas2007}
\bibitem{abdullatacas2007}
\bibinfo{author}{Parosh~Aziz \surnamestart Abdulla\surnameend},
  \bibinfo{author}{Giorgio \surnamestart Delzanno\surnameend},
  \bibinfo{author}{Noomene~Ben \surnamestart Henda\surnameend} \&
  \bibinfo{author}{Ahmed \surnamestart Rezine\surnameend}
  (\bibinfo{year}{2007}): \emph{\bibinfo{title}{Regular Model Checking Without
  Transducers (On Efficient Verification of Parameterized Systems)}}.
\newblock In: {\sl \bibinfo{booktitle}{Tools and Algorithms for the
  Construction and Analysis of Systems, 13th International Conference, {TACAS}
  2007, Held as Part of the Joint European Conferences on Theory and Practice
  of Software, {ETAPS} 2007 Braga, Portugal, March 24 - April 1, 2007,
  Proceedings}}, pp. \bibinfo{pages}{721--736},
  \doi{10.1007/978-3-540-71209-1_56}.

\bibitemdeclare{inproceedings}{undipcav2007}
\bibitem{undipcav2007}
\bibinfo{author}{Parosh~Aziz \surnamestart Abdulla\surnameend},
  \bibinfo{author}{Giorgio \surnamestart Delzanno\surnameend} \&
  \bibinfo{author}{Ahmed \surnamestart Rezine\surnameend}
  (\bibinfo{year}{2007}): \emph{\bibinfo{title}{Parameterized Verification of
  Infinite-State Processes with Global Conditions}}.
\newblock In: {\sl \bibinfo{booktitle}{Computer Aided Verification, 19th
  International Conference, {CAV} 2007, Berlin, Germany, July 3-7, 2007,
  Proceedings}}, pp. \bibinfo{pages}{145--157},
  \doi{10.1007/978-3-540-73368-3_17}.

\bibitemdeclare{article}{herdingcats}
\bibitem{herdingcats}
\bibinfo{author}{Jade \surnamestart Alglave\surnameend}, \bibinfo{author}{Luc
  \surnamestart Maranget\surnameend} \& \bibinfo{author}{Michael \surnamestart
  Tautschnig\surnameend} (\bibinfo{year}{2014}): \emph{\bibinfo{title}{Herding
  Cats: Modelling, Simulation, Testing, and Data Mining for Weak Memory}}.
\newblock {\sl \bibinfo{journal}{{ACM} Trans. Program. Lang. Syst.}}
  \bibinfo{volume}{36}(\bibinfo{number}{2}), pp. \bibinfo{pages}{7:1--7:74},
  \doi{10.1145/2627752}.

\bibitemdeclare{article}{AK86}
\bibitem{AK86}
\bibinfo{author}{Krzysztof~R. \surnamestart Apt\surnameend} \&
  \bibinfo{author}{Dexter \surnamestart Kozen\surnameend}
  (\bibinfo{year}{1986}): \emph{\bibinfo{title}{Limits for Automatic
  Verification of Finite-State Concurrent Systems}}.
\newblock {\sl \bibinfo{journal}{Inf. Process. Lett.}}
  \bibinfo{volume}{22}(\bibinfo{number}{6}), pp. \bibinfo{pages}{307--309},
  \doi{10.1016/0020-0190(86)90071-2}.

\bibitemdeclare{inproceedings}{parammc}
\bibitem{parammc}
\bibinfo{author}{Edmund~M. \surnamestart Clarke\surnameend},
  \bibinfo{author}{Orna \surnamestart Grumberg\surnameend} \&
  \bibinfo{author}{Michael~C. \surnamestart Browne\surnameend}
  (\bibinfo{year}{1986}): \emph{\bibinfo{title}{Reasoning About Networks With
  Many Identical Finite-State Processes}}.
\newblock In: {\sl \bibinfo{booktitle}{Proceedings of the Fifth Annual {ACM}
  Symposium on Principles of Distributed Computing, Calgary, Alberta, Canada,
  August 11-13, 1986}}, pp. \bibinfo{pages}{240--248},
  \doi{10.1145/10590.10611}.

\bibitemdeclare{inproceedings}{cubicletool}
\bibitem{cubicletool}
\bibinfo{author}{Sylvain \surnamestart Conchon\surnameend},
  \bibinfo{author}{Amit \surnamestart Goel\surnameend}, \bibinfo{author}{Sava
  \surnamestart Krstic\surnameend}, \bibinfo{author}{Alain \surnamestart
  Mebsout\surnameend} \& \bibinfo{author}{Fatiha \surnamestart
  Za{\"{\i}}di\surnameend} (\bibinfo{year}{2012}):
  \emph{\bibinfo{title}{Cubicle: {A} Parallel SMT-Based Model Checker for
  Parameterized Systems - Tool Paper}}.
\newblock In: {\sl \bibinfo{booktitle}{Computer Aided Verification - 24th
  International Conference, {CAV} 2012, Berkeley, CA, USA, July 7-13, 2012
  Proceedings}}, pp. \bibinfo{pages}{718--724},
  \doi{10.1007/978-3-642-31424-7_55}.

\bibitemdeclare{article}{German92}
\bibitem{German92}
\bibinfo{author}{Steven~M. \surnamestart German\surnameend} \&
  \bibinfo{author}{A.~Prasad \surnamestart Sistla\surnameend}
  (\bibinfo{year}{1992}): \emph{\bibinfo{title}{Reasoning about Systems with
  Many Processes}}.
\newblock {\sl \bibinfo{journal}{J. {ACM}}}
  \bibinfo{volume}{39}(\bibinfo{number}{3}), pp. \bibinfo{pages}{675--735},
  \doi{10.1145/146637.146681}.

\bibitemdeclare{inproceedings}{arraybased}
\bibitem{arraybased}
\bibinfo{author}{Silvio \surnamestart Ghilardi\surnameend},
  \bibinfo{author}{Enrica \surnamestart Nicolini\surnameend},
  \bibinfo{author}{Silvio \surnamestart Ranise\surnameend} \&
  \bibinfo{author}{Daniele \surnamestart Zucchelli\surnameend}
  (\bibinfo{year}{2008}): \emph{\bibinfo{title}{Towards {SMT} Model Checking of
  Array-Based Systems}}.
\newblock In: {\sl \bibinfo{booktitle}{Automated Reasoning, 4th International
  Joint Conference, {IJCAR} 2008, Sydney, Australia, August 12-15, 2008,
  Proceedings}}, pp. \bibinfo{pages}{67--82},
  \doi{10.1007/978-3-540-71070-7_6}.

\bibitemdeclare{inproceedings}{mcmt}
\bibitem{mcmt}
\bibinfo{author}{Silvio \surnamestart Ghilardi\surnameend} \&
  \bibinfo{author}{Silvio \surnamestart Ranise\surnameend}
  (\bibinfo{year}{2010}): \emph{\bibinfo{title}{{MCMT:} {A} Model Checker
  Modulo Theories}}.
\newblock In: {\sl \bibinfo{booktitle}{Automated Reasoning, 5th International
  Joint Conference, {IJCAR} 2010, Edinburgh, UK, July 16-19, 2010.
  Proceedings}}, pp. \bibinfo{pages}{22--29},
  \doi{10.1007/978-3-642-14203-1_3}.

\bibitemdeclare{article}{x86tsoshort}
\bibitem{x86tsoshort}
\bibinfo{author}{Peter \surnamestart Sewell\surnameend},
  \bibinfo{author}{Susmit \surnamestart Sarkar\surnameend},
  \bibinfo{author}{Scott \surnamestart Owens\surnameend},
  \bibinfo{author}{Francesco~Zappa \surnamestart Nardelli\surnameend} \&
  \bibinfo{author}{Magnus~O. \surnamestart Myreen\surnameend}
  (\bibinfo{year}{2010}): \emph{\bibinfo{title}{x86-TSO: a rigorous and usable
  programmer's model for x86 multiprocessors}}.
\newblock {\sl \bibinfo{journal}{Commun. {ACM}}}
  \bibinfo{volume}{53}(\bibinfo{number}{7}), pp. \bibinfo{pages}{89--97},
  \doi{10.1145/1785414.1785443}.

\end{thebibliography}

\end{document}